\pdfoutput=1
\documentclass[aps,prl,twocolumn,groupedaddress,showpacs,preprintnumbers]{revtex4-2}
\usepackage{epsfig}
\usepackage{graphics}
\usepackage{color}
\usepackage{calc}
\usepackage{hyperref}
\usepackage{array}
\usepackage{mathtools}
\usepackage{graphicx}
\usepackage[bitstream-charter, greekfamily=default]{mathdesign}
\usepackage{lipsum}
\usepackage{comment}
\usepackage{soul}

\usepackage{xcolor}

\newcommand{\de}{\mathrm{d}}

\usepackage{physics}
\usepackage{tikz}
\usetikzlibrary{calc,positioning}
\usepackage[capitalize]{cleveref}

\newsavebox{\panelbox}
\newcommand{\panel}[2]{\sbox{\panelbox}{#2}\usebox{\panelbox}\llap{\makebox[\wd\panelbox][l]{\raisebox{\dimexpr\ht\panelbox+2pt\relax}{\small (#1)}}}}

\begin{document}

\title{Corrections to the IPR in the localized phase of the Anderson model on the Bethe lattice}

\author{Tommaso Rizzo\textsuperscript{1,2}, Marco Tarzia\textsuperscript{3,4} and Angelica Vecchiarelli\textsuperscript{1,3}}

\affiliation{\textsuperscript{1}\mbox{Dip. Fisica, Universit\`a ``Sapienza'', Piazzale A. Moro 2, I-00185, Rome, Italy}\\
	\textsuperscript{2}\mbox{ISC-CNR, UOS Rome, Universit\`a ``Sapienza'', Piazzale A. Moro 2, I-00185, Rome, Italy}\\
	\textsuperscript{3} \mbox{LPTMC, CNRS-UMR 7600, Sorbonne Universit\'e, 4 Pl. Jussieu, F-75005 Paris, France}\\
	\textsuperscript{4} \mbox{Institut  Universitaire  de  France,  1  rue  Descartes,  75231  Paris  Cedex  05,  France}}

\begin{abstract}
We show that finite-size effects in the localized phase of the Anderson model on the Bethe lattice are highly anomalous. On random regular graphs of $N$ sites the inverse participation ratio (IPR) approaches its thermodynamic-limit value as $I_2(N)-I_2^\infty\sim N^{-\kappa(W)}$, with a disorder-dependent exponent $0<\kappa<1$ throughout the localized phase, and the same exponent governs the correction due to a finite imaginary regulator, $I_2(\eta)-I_2^\infty\sim\eta^{\kappa(W)}$. We derive $\kappa = 2\beta-1$, with $\beta$ the power-law tail exponent of the local density of states distribution, in a solvable model sharing the criticality of the Anderson model on the Bethe lattice. We then show that the same mechanism operates in the Anderson model itself, and find quantitative agreement with the numerics. A self-consistent one-loop treatment of the loops of finite graphs independently reproduces the finite-size corrections. Physically, a spectral identity and an exact sum rule tie both corrections to rare Mott-like resonances hybridizing distant localized eigenstates: a finite graph lacks the resonant pairs whose centers would lie farther apart than its size allows. Our results identify the scaling variable needed to extrapolate IPR data to the thermodynamic limit.
\end{abstract}

\maketitle

Anderson localization (AL)~\cite{anderson1958absence} is one of the
most spectacular phenomena in condensed matter
physics~\cite{lee1985disordered,evers2008anderson,lagendijk2009fifty}.
AL on the Bethe lattice (BL) occupies a distinguished position: the
Anderson model is exactly solvable there~\cite{anderson1973selfconsistent,efetov1985anderson,efetov1987density,efetov1987anderson,zirnbauer1986localization,zirnbauer1986anderson,verbaarschot1988graded,mirlin1991localization,mirlin1991universality,fyodorov1991localization,fyodorov1992novel,mirlin1994statistical,mirlin1994distribution,tikhonov2019statistics,tikhonov2019critical,biroli2010anderson,parisi2019anderson,pascazio2023anderson,rizzo2024localized,bandeira2026andersonlocalizationbethelattice};
it provides a single-particle analogue of ergodicity breaking and many-body localization (MBL)~\cite{basko2006metal,gornyi2005interacting,altshuler1997quasiparticle,altman2015universal,nandkishore2015many,abanin2017recent,alet2018many,abanin2019colloquium,sierant2024many},
since the dynamics of an interacting system is a hopping problem on the high-dimensional graph of its Hilbert space; and, since the upper critical dimension of AL is $d_u=\infty$~\cite{tarquini2017critical,castellani1986upper,mirlin1994distribution,baroni2024corrections},
it is the natural mean-field starting point for localization both in
real and in Hilbert space.

Despite its exact solution, the BL model continues to reveal
subtleties~\cite{tikhonov2019statistics,tikhonov2019critical,biroli2010anderson,parisi2019anderson,rizzo2024localized,bandeira2026andersonlocalizationbethelattice}:
its numerical study suffers from finite-size effects so strong that they
have fueled a long-standing controversy over the validity of the
asymptotic solution itself~\cite{biroli2012difference,de2013ergodicity,kravtsov2018non,pino2020scaling,bera2018return,de2020subdiffusion,tikhonov2016anderson,garcia2017scaling,biroli2022critical,de2014anderson,altshuler2016nonergodic,garcia2020two,tikhonov2021anderson}.

We focus on the finite-size behavior of the inverse participation ratio
(IPR), $I_2 = \langle \sum_i |\psi_\alpha(i)|^4 \rangle$, the order
parameter of the localized phase. $I_2$ vanishes as $1/N$ in the
delocalized phase, jumps discontinuously to a finite value at the
transition~\cite{efetov1985anderson,efetov1987density,efetov1987anderson,zirnbauer1986localization,zirnbauer1986anderson,verbaarschot1988graded,mirlin1991localization,mirlin1991universality,fyodorov1991localization,fyodorov1992novel,mirlin1994statistical}, and then grows with a square-root
singularity in the localized phase~\cite{rizzo2024localized}. Extracting
its thermodynamic limit from finite-size numerics is therefore delicate,
and the correct finite-size scaling (FSS) of the IPR is still
debated~\cite{tikhonov2019statistics,garcia2022critical,10.21468/SciPostPhys.15.2.045,garcia2017scaling,chen2023quantum,chen2024describing,vanoni2023renormalization,garcia2026connecting}.

Building on the precise control of the asymptotic IPR achieved in
Ref.~\cite{rizzo2024localized}, we show that in the localized phase the
corrections to the IPR decay with an unusual, continuously varying
exponent $\kappa(W) \in (0,1)$, even far from $W_c$, and that the same
exponent controls the finite-size correction and the correction due to a
finite imaginary regulator $\eta$. Using a
solvable toy non-linear $\sigma$-model (NL$\sigma$M) on the hyperbolic
superplane~\cite{zirnbauer1990fourier,zirnbauer1991fourier,Gruzberg_1996,dupre1996localization},
which has the same unusual critical behavior as the Anderson model on the
BL, we derive $\kappa = 2\beta - 1$, with $\beta$ the tail exponent of the
local density of states (LDoS) distribution~\cite{companion}. We then
show that the same mechanism operates in the cavity equations of the
Anderson model under a few explicit assumptions, and that the prediction
agrees with the numerics with no adjustable parameter. Finally, a
self-consistent one-loop treatment of the cavity equations reproduces the
finite-size corrections quantitatively.

These results have a transparent physical interpretation. A spectral
identity and an exact sum rule tie both corrections to the same function $N^2\mathcal{C}(\omega)$, proportional to the average overlap of two eigenstates at energy distance $\omega$, on which $\eta$ and $N$ act as infrared cutoffs. Two
eigenstates close in energy are typically localized far apart and overlap
exponentially little, but the average overlap is dominated by rare pairs
with an overlap of order one. As argued in
Ref.~\cite{tikhonov2021eigenstate}, these are Mott-like resonances: two
localized states $\psi_A$ and $\psi_B$ with distant centers hybridize into
$\psi_1 = \cos\theta\, \psi_A + \sin\theta\, \psi_B$ and $\psi_2 =
\sin\theta\, \psi_A - \cos\theta\, \psi_B$, which, unlike $\psi_A$ and
$\psi_B$, have a finite overlap $\simeq \sin^2\theta\cos^2\theta\,(I_A +
I_B)$ and a reduced IPR, $\simeq \cos^4\theta\, I_A + \sin^4\theta\, I_B$
for $\psi_1$, with $I_{A,B}$ the IPRs of $\psi_{A,B}$. Through the spectral
identity, $\kappa = 2\beta-1$ implies that the density of such pairs
behaves as $\omega^{-2(1-\beta)}$ at small splitting $\omega$, which a
Mott-type counting based on the exact statistics of the eigenstate
amplitudes accounts for. A given eigenstate therefore has a small
probability of belonging to such a hybridized, doubly-peaked pair. On a finite graph the pairs
whose two centers would lie farther apart than the typical distance
$\sim\ln N/\ln k$ cannot form, and exact diagonalization shows that this
deficit accounts for most of the finite-size correction.

The weight of the resonances with splittings below $\omega$, $\propto
\omega^{2\beta-1}$, vanishes as $\omega \to 0$ only for $\beta > 1/2$:
hybridized states become less and less rare as $W_c$ is approached, where
$\beta \to 1/2$~\cite{anderson1973selfconsistent}, and localization is
lost when resonances
proliferate~\cite{anderson1958absence,abou1974self,aizenman2013resonant}.
On the delocalized side, eigenstates are built from the hybridization of
many localized states whose distance diverges at the transition, as in the
picture of Ref.~\cite{mirlin1994distribution}.

We consider the standard Anderson tight-binding Hamiltonian
\begin{equation} \label{eq:H}
    \mathcal{H} = -t \sum_{\langle i,j \rangle} \big( |i\rangle\langle j| + |j\rangle\langle i| \big)
    + \sum_{i=1}^N \epsilon_i |i\rangle\langle i| \, ,
\end{equation}
with $t=1$ and independent on-site energies $\epsilon_i$ uniform in
$[-W/2,W/2]$. The BL is realized at finite $N$ by $(k+1)$-regular random
graphs (RRG), which are locally tree-like up to loops of typical length
$\sim \ln N/\ln k$~\cite{wormald1999models,marinari2004circuits,marinari2006number}. We
fix $k=2$ throughout, for which the localization transition sits at
$W_c \simeq 18.17 \pm 0.01$~\cite{tikhonov2019critical,parisi2019anderson}.

The IPR $I_2$, averaged over eigenstates $\alpha$ at
the band center $E_\alpha=0$ and over disorder, has an asymptotic value
$I_2^\infty(W)$ that follows from the resolvent $G(z) = [\mathcal{H}-z]^{-1}$,
$z=E+{\rm i}\eta$, whose diagonal elements $G_{ii}$ (the local Green's
functions) determine the IPR exactly~\cite{mirlin2000statistics,rizzo2024localized}:
\begin{equation} \label{eq:I2G}
    I_2(\eta) = \frac{\eta \langle |G_{ii}|^2 \rangle}{\langle {\rm Im}\,
    G_{ii} \rangle} \, ,
\end{equation}
with $I_2^\infty = \lim_{\eta \to 0^+} I_2(\eta)$ at $N=\infty$. On the
infinite BL, $G_{ii}$ follows exactly from the ``cavity'' Green's functions
$G_{i\to j}$, the diagonal elements of the resolvent with the neighbor $j$
of $i$ removed, which obey the closed recursion~\cite{anderson1973selfconsistent,abou1974self}
\begin{equation} \label{eq:cavity}
    G_{i \to j}^{-1} = \epsilon_i - {\rm i}\eta - \!\! \sum_{m\in\partial
    i \setminus j} \!\! G_{m \to i} \, ,
\end{equation}
from which the local Green's function follows as
\begin{equation} \label{eq:Gii}
    G_{ii}^{-1} = \epsilon_i - {\rm i}\eta - \sum_{m\in\partial i} G_{m
    \to i} \, .
\end{equation}
Solving Eq.~\eqref{eq:cavity} by population dynamics~\cite{anderson1973selfconsistent,mezard2001bethe} on the infinite BL gives $I_2(\eta)$, hence $I_2^\infty$ in the limit $\eta \to 0^+$~\cite{rizzo2024localized} and the finite-regulator correction $I_2(\eta)-I_2^\infty$; exact diagonalization of finite RRGs at $\eta=0$ gives the finite-size correction
$I_2(N)-I_2^\infty$~\footnote{The $O(1/N)$ bias due to the absence of an
eigenvalue exactly at $E = 0$ is removed by reweighting the eigenstates,
exactly up to an $O(1/N)$ change of the disorder distribution.}.

\begin{figure*}[t]
\centering
\panel{a}{\includegraphics[width=0.325\textwidth]{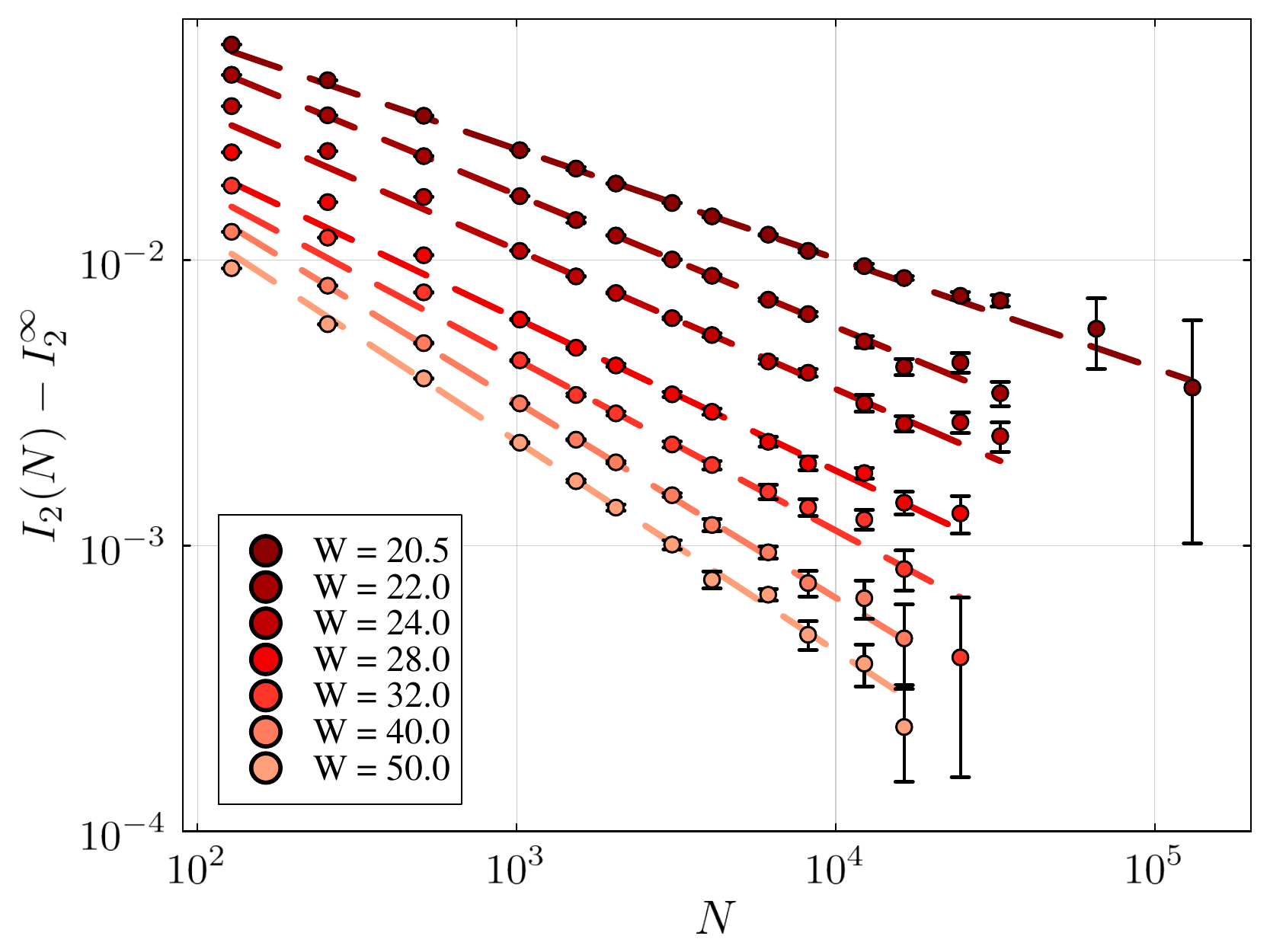}} \hfill
\panel{b}{\includegraphics[width=0.325\textwidth]{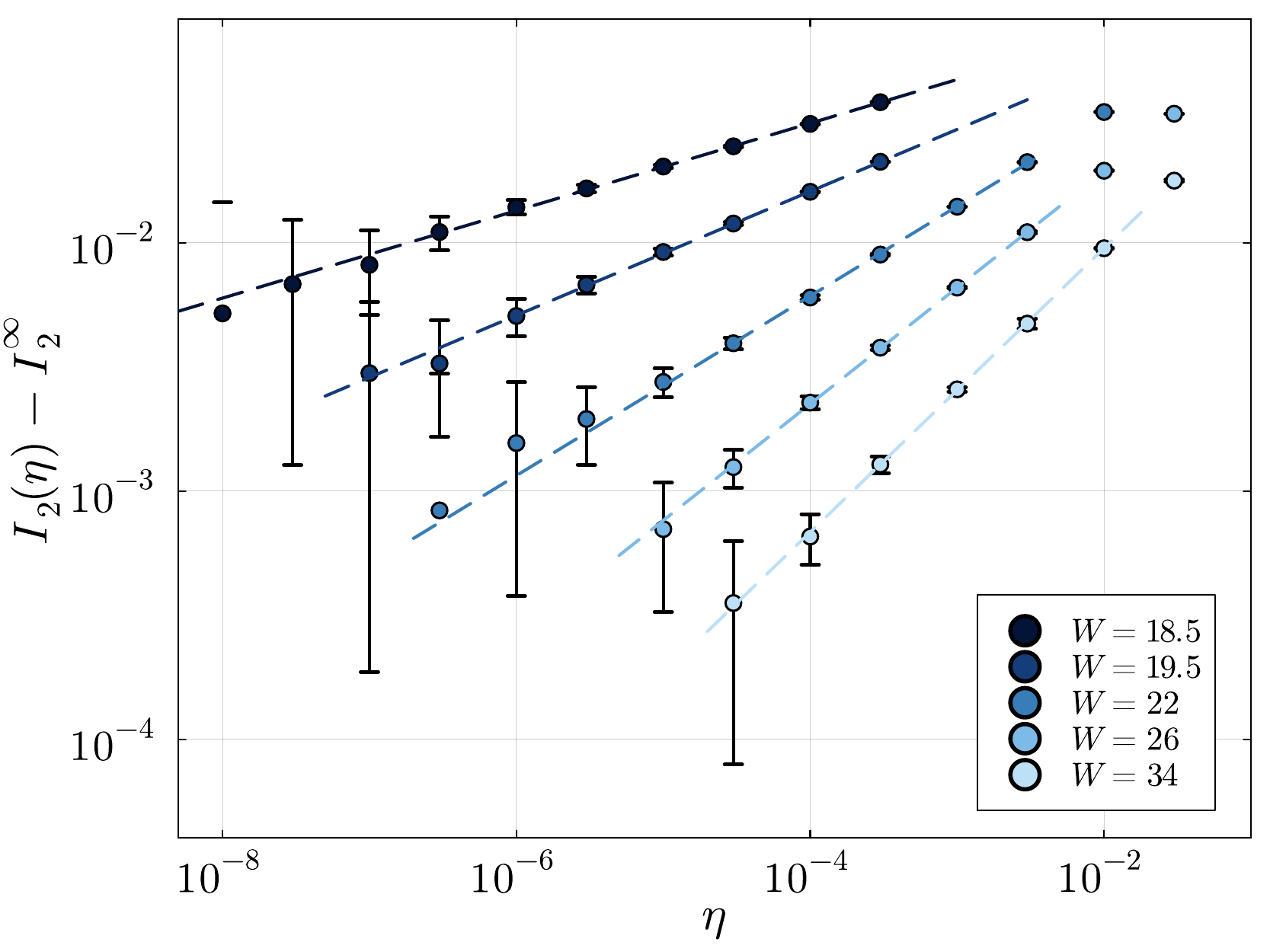}} \hfill
\panel{c}{\includegraphics[width=0.325\textwidth]{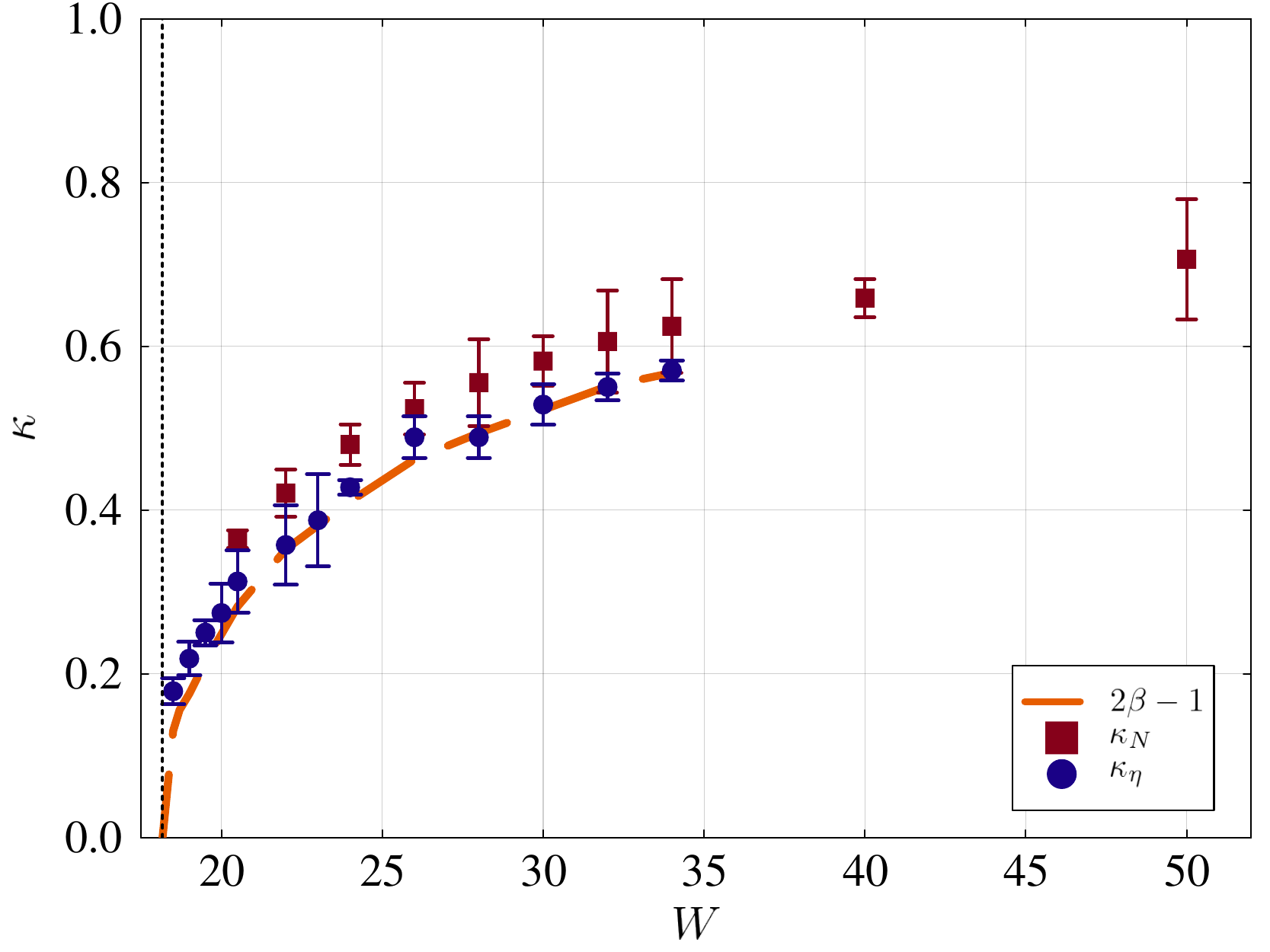}}
\caption{Corrections to the asymptotic IPR in the localized phase, for
several $W$. (a) $I_2(N) - I_2^\infty$ vs.\ $N$ at $\eta=0$, from exact
diagonalization; (b) $I_2(\eta) - I_2^\infty$ vs.\ $\eta$ at $N=\infty$,
from the cavity solution. Dashed lines: power-law fits to
Eqs.~\eqref{eq:corrN} and~\eqref{eq:correta}. (c) Exponents $\kappa_N$
(squares) and $\kappa_\eta$ (circles) from these fits vs.\ $W$. Dashed line:
parameter-free prediction $\kappa = 2\beta(W) - 1$
[Eq.~\eqref{eq:toyresult}], with $\beta(W)$ obtained independently from
Eq.~\eqref{eq:betatail}.}
\label{fig:main}
\end{figure*}

Naively, both corrections would be expected to be regular. Instead, we find that \emph{throughout the localized phase} both corrections follow power laws [Figs.~\ref{fig:main}(a) and (b)],
\begin{subequations} \label{eq:corrections}
\begin{align}
    I_2(N,\eta{=}0) &\simeq I_2^\infty + c_N \, N^{-\kappa_N(W)} \, ,
    \label{eq:corrN}\\
    I_2(N{\to}\infty,\eta) &\simeq I_2^\infty + c_\eta \, \eta^{\kappa_\eta(W)} \, ,
    \label{eq:correta}
\end{align}
\end{subequations}
with exponents $\kappa_N, \kappa_\eta \in (0,1)$ that vary
\emph{continuously} with the disorder strength $W$, increasing
monotonically as $W$ is raised away from $W_c$ [Fig.~\ref{fig:main}(c)]. Such a non-universal exponent throughout an entire
phase is reminiscent of the low-temperature phase of the
Berezinskii--Kosterlitz--Thouless transition~\cite{berezinskii1971destruction,kosterlitz1973ordering},
to which the Anderson transition on RRGs has been
compared~\cite{garcia2022critical,vanoni2023renormalization,garcia2026connecting}, and more closely
of Griffiths phases~\cite{griffiths1969nonanalytic,vojta2006rare},
where power laws arise from rare regions: here, as we show below, $\kappa$
is inherited from the power-law tail of the LDoS
distribution, i.e., from rare resonant configurations --- to our knowledge
the first instance of such a mechanism governing the finite-size
corrections of an order parameter far from criticality. The prefactors, by contrast, depend only weakly on $W$ deep in the localized phase~\footnote{For $k=2$, $c_\eta \approx 0.15$ at $W=22$ and $0.13$ at $W=34$, while
fits with $\kappa = 2\beta-1$ give $c_N = 0.15 \pm 0.02$ at $W=22$ and
$c_N = 0.18 \pm 0.02$ at $W=34$.}. Both power laws are asymptotic: they hold for $\kappa\ln N \gg 1$ [$\kappa\ln(1/\eta) \gg 1$], a regime that moves to ever larger sizes as $W \to W_c$, where the prefactors vanish $\propto 2\beta-1$ and the corrections cross over to the critical $1/\ln N$ behavior~\cite{companion}.

Another striking feature is that the two exponents agree, $\kappa_N(W)
\simeq \kappa_\eta(W) \equiv \kappa(W)$, within the accuracy of the fits at all the disorder strengths probed [Fig.~\ref{fig:main}(c)]: a genuine finite-size effect and an
artificial spectral broadening share a common exponent. As we show below, $\kappa = 2\beta - 1$ is fixed by the
tail exponent $\beta$ of the LDoS distribution; this parameter-free
prediction [dashed line in Fig.~\ref{fig:main}(c)] tracks $\kappa_N$ and
$\kappa_\eta$ across the entire localized phase~\footnote{Close to $W_c$ the power-law fits overestimate $2\beta-1$, since $\kappa\ln N$ is not large; the small excess of $\kappa_N$ over $\kappa_\eta$, visible at all $W$, decreases slowly with $N$.}.

To see why the two corrections share the same exponent, insert the spectral form $G_{ii} =
\sum_\alpha |\psi_\alpha(i)|^2/(E_\alpha - {\rm i}\eta)$ into the
numerator of Eq.~\eqref{eq:I2G}. $I_2(\eta)$ then separates into a
diagonal piece, the IPR of the $O(N\eta)$ eigenstates within $\sim\eta$ of
$E=0$, which differs from $I_2^\infty$ only by analytic $O(\eta)$ terms,
and an off-diagonal piece, in which pairs of eigenstates $\alpha \neq
\alpha'$ close in energy enter with a weight $O(1/\eta)$ times their spatial
overlap. Averaged over a
window of energies much wider than $\eta$, the latter gives
\begin{equation} \label{eq:spectralpair}
    I_2(\eta) - I_2^\infty \simeq \langle\rho\rangle \int_{-\infty}^{\infty} {\rm d}\omega \, N^2\mathcal{C}(\omega)\,
    \frac{4\eta^2}{\omega^2 + 4\eta^2} \, ,
\end{equation}
where $\langle\rho\rangle$ is the average density of states per site and
$N^2\mathcal{C}(\omega) = N \langle \sum_i |\psi_\alpha(i)|^2|\psi_{\alpha'}(i)|^2 \rangle$,
averaged over the pairs of eigenstates with energies $E_\alpha = 0$ (the
reference energy) and $E_{\alpha'} = \omega$, is the eigenstate correlation function of
Refs.~\cite{tikhonov2019statistics,tikhonov2021eigenstate} [denoted $\beta(\omega)$ there].
Its finite-size counterpart follows from an exact sum rule: since
$\sum_{\alpha'} |\psi_{\alpha'}(i)|^2 = 1$, the IPR of an eigenstate is $1 -
\sum_{\alpha' \neq \alpha} \sum_i |\psi_\alpha(i)|^2|\psi_{\alpha'}(i)|^2$,
one minus its total overlap with all the others. At $\eta = 0$ the finite-size correction is therefore the deficit
of this total overlap on a graph of $N$ sites,
\begin{equation} \label{eq:sumrule}
    I_2(N) - I_2^\infty \simeq \langle\rho\rangle \int_{-\infty}^{\infty} {\rm d}\omega \,
    \big[ N^2\mathcal{C}_\infty(\omega) - N^2\mathcal{C}_N(\omega) \big] \, ,
\end{equation}
with $N^2\mathcal{C}_\infty \equiv \lim_{N\to\infty} N^2\mathcal{C}_N$ the function of Eq.~\eqref{eq:spectralpair}, the deficit being concentrated at small $|\omega|$~\cite{companion}:
a finite $\eta$ does not resolve the pairs with splittings $\omega \lesssim
\eta$, while a finite graph lacks them. Two states within $\eta$ in energy
are typically at distance $d_{\rm typ} \simeq \ln(1/\eta)/\ln k$ and overlap
exponentially little, but the rare resonant
pairs~\cite{mott1968conduction,cuevas2007two,ivanov2012hybridization} that
dominate the average overlap and its higher moments give
$N^2\mathcal{C}(\omega) \propto |\omega|^{-\mu(W)}$, $0 < \mu <
1$~\cite{tikhonov2021eigenstate}; inserted in Eq.~\eqref{eq:spectralpair},
this gives $I_2(\eta) - I_2^\infty \propto \eta^{1-\mu}$, i.e.\ $\kappa =
1 - \mu$. In Eq.~\eqref{eq:sumrule}, exact diagonalization shows that
the deficit builds up at splittings below a few mean level spacings, $\omega_N \sim 1/(N\langle\rho\rangle)$~\cite{companion},
so that, if it scales with $\omega_N$, $I_2(N) - I_2^\infty \propto \omega_N^{1-\mu} \propto N^{-\kappa}$
with the same exponent. Equations~\eqref{eq:corrections} then suggest a single scaling form, of
which they are the two opposite limits,
\begin{equation} \label{eq:scaling}
    I_2(N,\eta) - I_2^\infty = N^{-\kappa(W)}\, \Psi(N\eta) \, ,
\end{equation}
with $\Psi(y) \to c_N$ as $y \to 0$ and $\Psi(y) \to c_\eta\, y^{\kappa}$
as $y \to \infty$, in the asymptotic regime $\kappa\ln N, \kappa\ln(1/\eta) \gg 1$: two ways of imposing the same infrared cutoff on the
eigenstate correlations.

The exponent governing Eqs.~\eqref{eq:corrections} and~\eqref{eq:scaling} is tied to a
property of the exact solution of the Anderson model on the BL: in the
localized phase, the joint distribution $Q_{0^+}(m,\hat m)$ of the real part $m$ and the rescaled imaginary part $\hat m$ of the inverse local Green's function
$M_{ii} \equiv G_{ii}^{-1} = m - {\rm i}\eta\hat m$, evaluated in the
$\eta \to 0^+$ limit, develops a power-law tail at $m=0$:
\begin{equation} \label{eq:betatail}
    Q_{0^+}(0,\hat m) \simeq \text{cst} \, \hat m^{-1-\beta} \, ,
    \qquad \hat m \to \infty \, ,
\end{equation}
with a disorder-dependent exponent
$\beta \simeq 1/2 + c_\beta\sqrt{W-W_c}$ near
$W_c$~\cite{efetov1985anderson,efetov1987density,efetov1987anderson,zirnbauer1986localization,zirnbauer1986anderson,verbaarschot1988graded,mirlin1991localization,mirlin1991universality,fyodorov1991localization,fyodorov1992novel,mirlin1994statistical,mirlin1994distribution,tikhonov2019statistics,rizzo2024localized}.
At a resonant site $\hat m^{-1}$ is the LDoS in units of $1/\eta$, i.e.\ the
weight $|\psi_\alpha(i)|^2$ of the resonant eigenstate on that
site~\cite{mirlin1994distribution,rizzo2024localized}.

This gives $\beta$ a real-space meaning. A large $\hat m$ signals an
eigenstate localized far away, whose tail reaches the site with a small
weight $u = 1/\hat m$, distributed as $T_\infty(u) \propto
u^{\beta-2}$~\cite{mirlin1994distribution} (exact diagonalization confirms
that the tail is carried by a single distant eigenstate~\cite{companion}).
A phenomenological, Mott-type counting~\cite{mott1968conduction,tikhonov2021eigenstate}
then fixes $\mu$. A state localized at the site hybridizes with such a
distant state when their energy separation is smaller than their coupling
$\sim \sqrt u$, giving a pair with overlap $O(1)$ and splitting $\omega
\sim \sqrt u$; the density of such pairs, $\propto T_\infty(u)\sqrt
u\,{\rm d}u$ at $u \sim \omega^2$, gives $N^2\mathcal{C} \propto
\omega^{-2(1-\beta)}$, i.e.\ $\kappa = 1 - \mu = 2\beta - 1$. This is the
estimate of Ref.~\cite{tikhonov2021eigenstate}, with the average decay of
the wavefunctions replaced by their exact statistics: the rare large
overlaps are controlled by the same small-$u$ tail that makes the typical
overlap exponentially small. Exact
diagonalization~\cite{tikhonov2021eigenstate,companion} and population
dynamics directly on the infinite BL at $\eta = 0^+$~\cite{companion} confirm $\mu = 2(1-\beta)$, with the amplitude
fixed by Eq.~\eqref{eq:spectralpair}; the generalized overlaps
$N\langle\sum_i |\psi_\alpha(i)|^{2p}|\psi_{\alpha'}(i)|^{2s}\rangle(\omega)$
all decay with the same exponent, so that all the integer moments $I_q = \langle\sum_i|\psi_\alpha(i)|^{2q}\rangle$ (more generally those with $q > \beta$) share $\kappa$, and the two eigenstates of a hybridized pair are spread over the
same two centers, at a distance growing as
$\ln(1/\omega)$~\cite{companion}.

To see analytically how $\beta$ propagates into the finite-$\eta$
corrections, we use a solvable ``toy'' NL$\sigma$M on the hyperbolic
superplane~\cite{zirnbauer1990fourier,zirnbauer1991fourier,disertori2010quasi,disertori2010anderson,sabot2015edge}, which shares
the non-compact graded symmetry of the true Anderson transition and, as
shown by Gruzberg and Mirlin~\cite{Gruzberg_1996} (see also
Ref.~\cite{dupre1996localization}), its exotic BL criticality, including
the structure of the exponent $\beta$.

The toy model lives on five-component supervectors $\overline{\psi} = (\sigma,
\pi_1,\pi_2,\overline\xi,-\xi)$ of unit hyperbolic norm, $\vert \vert \psi \vert \vert^2 = \sigma^2 - \pi_1^2 - \pi_2^2 - 2 \overline{\xi} \xi = 1$, with Hamiltonian
$H_{\rm hs} = J\sum_{\langle ij\rangle} \overline\psi_i \cdot \psi_j + H
\sum_i \sigma_i$ on a BL of branching ratio $k$. Here $J$ plays the role
of $1/W$, and the symmetry-breaking field $H$ that of $\eta$, regularizing
the same non-compact zero mode. Its paramagnetic phase, $J < J_c$, on which we focus, is the
counterpart of the localized phase. The model is solved by a cavity
recursion for the field distribution $\tilde
P(\theta)$~\cite{Gruzberg_1996,companion}, whose IPR proxy is $I_2(H) = e^{-2H/(k-1)}\, H \int_0^\infty
\!\de\theta\, e^{-H\cosh\theta} \tilde P^{k+1}(\theta) \sinh\theta$.

With $\cosh \theta=e^t/H$ and $P_H(t) \equiv \tilde{P}(\theta)$, the $H=0$
cavity equation linearized around its trivial fixed point $P \equiv 1$
gives $1-P_0(t) \propto e^{\beta t}$, where $\beta$ solves
$1/k = K_{\beta-1/2}(J)/K_{-1/2}(J)$~\cite{Gruzberg_1996} --- the counterpart of $\beta$ in Eq.~\eqref{eq:betatail}, fixed by a condition with the same structure as the Abou-Chacra--Anderson--Thouless one.
Since $K_\nu = K_{-\nu}$, this equation has two roots, $\beta$ and
$1-\beta$, merging at $1/2$ when $J \to J_c$; the $H=0$ solution is
controlled by the larger root $\beta>1/2$.
A small $H$ perturbs this fixed point in two regimes: a bulk regime at
$t=O(1)$, where the correction is the solution $f_{\rm typ}(t)$ of a linear
integral equation, and a boundary layer at $t - \ln H = O(1)$. Matching
them~\cite{companion} gives
\begin{equation} \label{eq:atyp}
    P_H(t) - P_0(t) = a_{\rm typ}(H)\, f_{\rm typ}(t) \, , \qquad
    a_{\rm typ}(H) \propto H^{2\beta-1} \, ,
\end{equation}
because the perturbation of weight $\propto H^{\beta}$ created in the
boundary layer propagates into the bulk along the subdominant root,
$f_{\rm typ}(t) \simeq e^{(1-\beta)t}$ for $t \to -\infty$, which amplifies
it by $H^{-(1-\beta)}$. Since the
IPR proxy $I_2(H)$ is an integral of $P_H(t)$ dominated by the bulk
region, Eq.~\eqref{eq:atyp} directly yields
\begin{equation} \label{eq:toyresult}
    I_2(H) - I_2(0) \propto H^{2\beta - 1} \, ,
\end{equation}
i.e., $\kappa = 2\beta - 1$, the relation found numerically in
Fig.~\ref{fig:main}(c). At the critical point the two roots coalesce and
the matching gives a logarithmic correction, $I_2(H)-I_2(0) \propto
1/|\ln H|$, reached only for $|\ln H| \gtrsim 50$, so that the effective
powers of $\ln(1/\eta)$ fitted in Ref.~\cite{tikhonov2019statistics} are
preasymptotic~\cite{companion}. Close to $J_c$ the matching yields the whole crossover, $I_2(H)-I_2(0) \propto \kappa/(H^{-\kappa}-1)$, from $\kappa H^{\kappa}$ for $\kappa|\ln H| \gg 1$, with a prefactor vanishing linearly in $\kappa$, to $1/|\ln H|$ at $J_c$~\cite{companion}. Since $N$ plays the role of $1/\eta$ [Eq.~\eqref{eq:scaling}], we correspondingly expect $I_2(N) - I_2^\infty \propto \kappa/(N^{\kappa}-1)$ close to $W_c$, i.e.\ $1/\ln N$ times a function of the scaling variable $\kappa\ln N$, which becomes $\propto 1/\ln N$ at $W_c$; exact diagonalization and population dynamics support this form~\cite{companion}.
The same mechanism operates in the cavity equations~\eqref{eq:cavity} of the Anderson model, under a few explicit assumptions supported by the numerics~\cite{companion}. Written for the joint distribution of the real and rescaled imaginary parts of the cavity Green's functions, these equations depend on $\eta$ only through $\eta^2$, so that a non-analytic correction to this distribution can only arise from a zero mode of the $\eta=0$ recursion linearized around its fixed point. Its tail $\propto \hat m^{-(2-\beta)}$ is governed by the second root $1-\beta$ of the Abou-Chacra--Anderson--Thouless condition~\cite{anderson1973selfconsistent,abou1974self}, and matching it to the boundary layer at $\hat m \sim 1/\eta$ gives again an amplitude $\propto \eta^{2\beta-1}$. Beyond the boundary layer the distribution decays as $\hat m^{\beta-3}$~\cite{companion}.

\begin{figure*}[t]
\centering
\panel{a}{\includegraphics[width=0.46\textwidth]{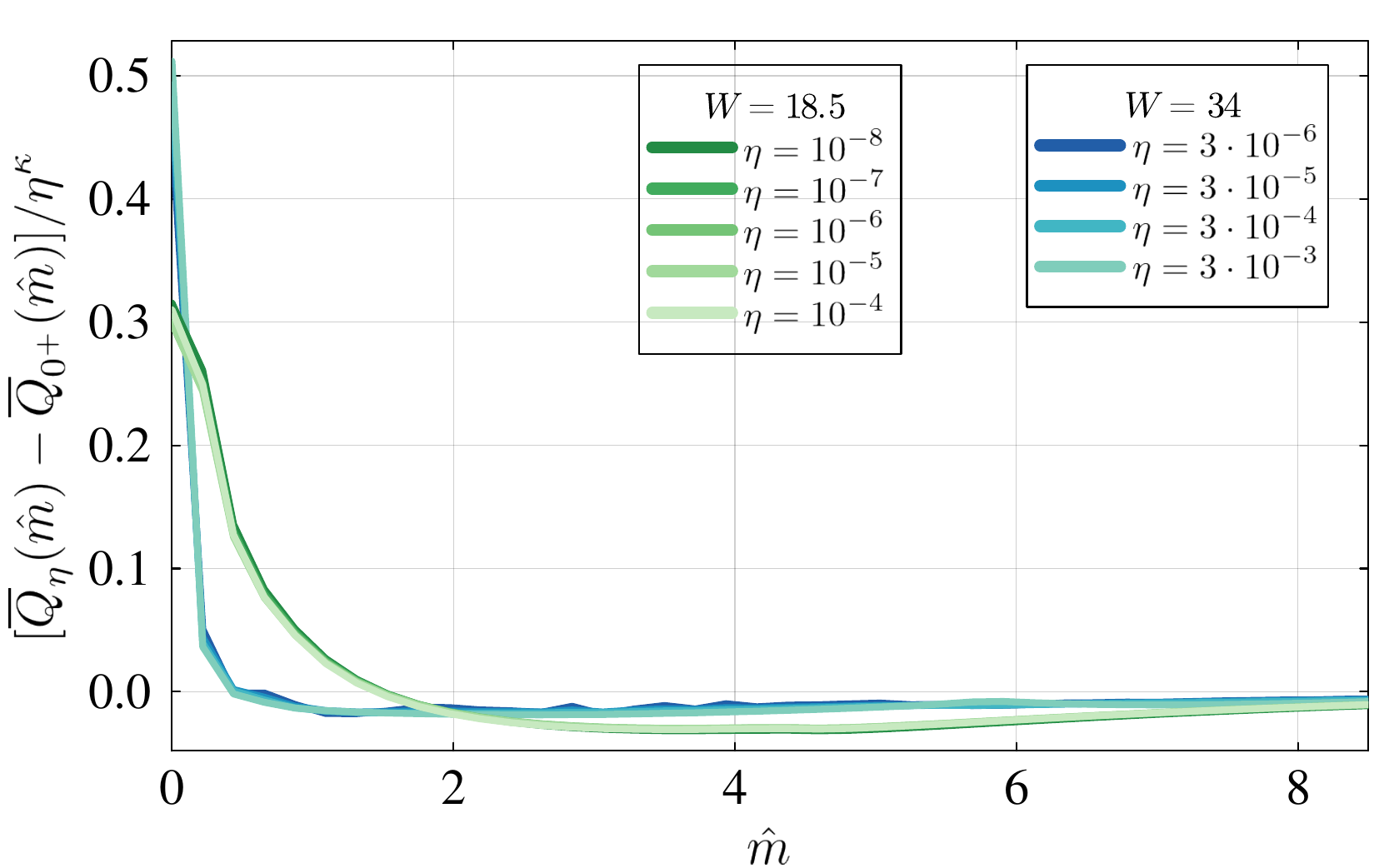}} \hfill
\panel{b}{\includegraphics[width=0.46\textwidth]{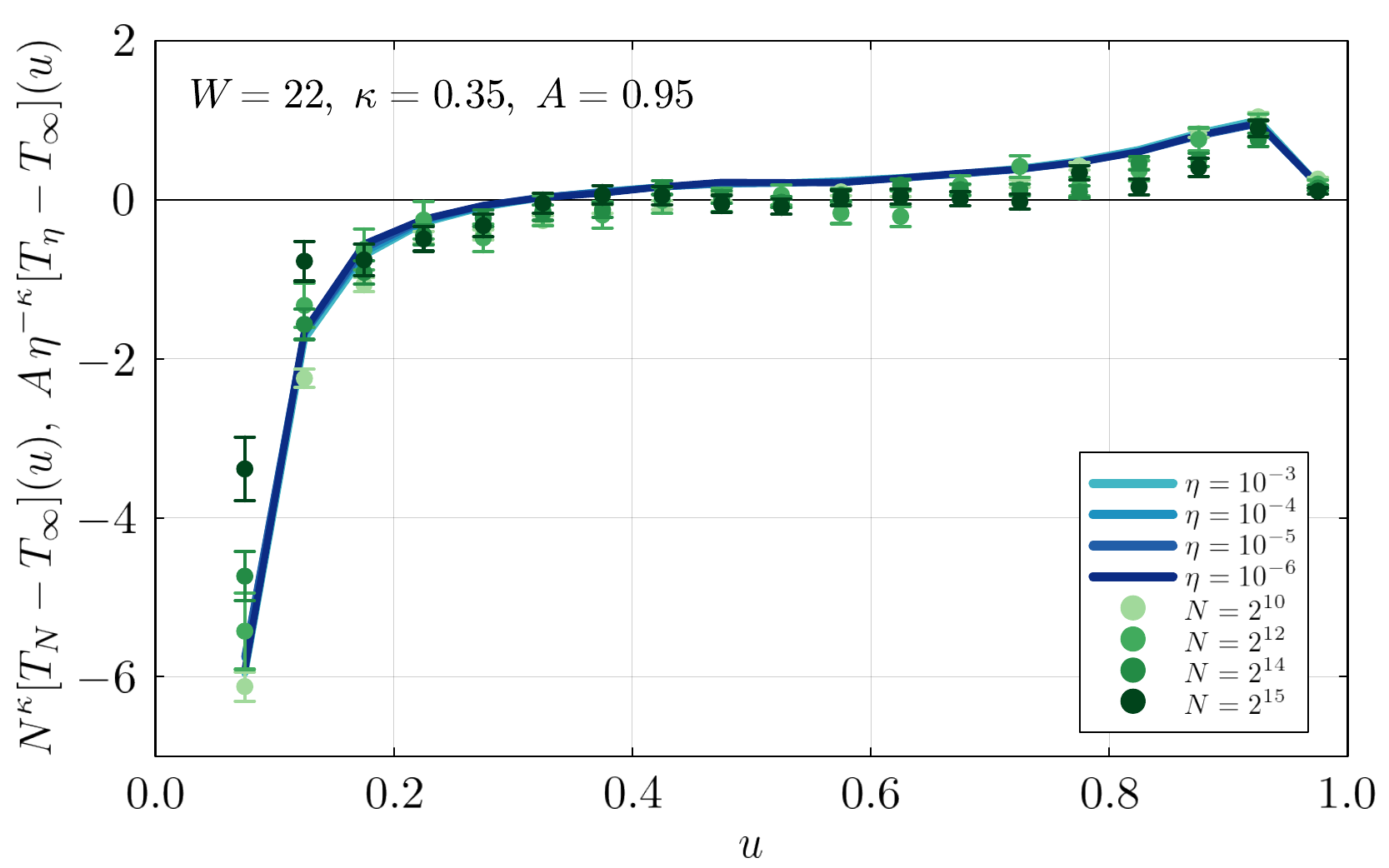}}
\caption{Scaling collapse of the full distributions underlying the IPR,
in the bulk region $\hat m, u = O(1)$, with $\kappa = 2\beta(W) - 1$. (a) $[\overline{Q}_\eta(\hat m) -
\overline{Q}_{0^+}(\hat m)]/\eta^{\kappa}$ for the local-Green's-function
variable $\hat m$ [Eq.~\eqref{eq:betatail}], for several $\eta$ at
$W = 18.5$ and $34$.
(b) $[T_N(u) - T_\infty(u)]\,N^{\kappa}$ for the eigenstate amplitudes
$u = |\psi_\alpha(i)|^2$, from exact diagonalization for several $N$
(symbols), compared with the finite-$\eta$ master curve
$A\,\eta^{-\kappa}[T_\eta(u) - T_\infty(u)]$ with $T_\eta(u) \equiv
u^{-3}\overline{Q}_\eta(1/u)$, from population dynamics for several $\eta$
(lines), at $W = 22$, with $A = 0.95$.}
\label{fig:collapse}
\end{figure*}

The toy model predicts more than the scaling of the IPR: by
Eq.~\eqref{eq:atyp}, the \emph{entire} correction to the bulk
distribution factorizes into an amplitude $\propto H^{2\beta-1}$ times an
$H$-independent function $f_{\rm typ}(t)$. The same should hold for the
full distributions of the Anderson problem.

We test this directly on the two distributions playing the role of
$P_H(t)$ in the finite-$\eta$ and finite-$N$ problems: the
distribution $\overline{Q}_\eta(\hat m) = Q_\eta(0,\hat m) / \int_0^\infty Q_{0^+}(0,\hat m) \, \de \hat m$ of the local-Green's-function
variable $\hat m$ at $m=0$, and the distribution $T_N(u)$ of the eigenstate
amplitudes $u = |\psi_\alpha(i)|^2$ at finite $N$, related for $N \to \infty$ by $T_\infty(u) =
\overline{Q}_{0^+}(1/u)\,u^{-3}$~\cite{mirlin1994distribution,rizzo2024localized}. In direct analogy with
Eq.~\eqref{eq:atyp}, we conjecture
\begin{subequations} \label{eq:distscaling}
\begin{align}
    \overline{Q}_\eta(\hat m) - \overline{Q}_{0^+}(\hat m) &= \eta^{\kappa} \,
    g_\eta(\hat m) \, , \label{eq:distscalingQ}\\
    T_N(u) - T_\infty(u) &= N^{-\kappa} \, g_N(u) \, , \label{eq:distscalingT}
\end{align}
\end{subequations}
with the \emph{same} exponent $\kappa = 2\beta-1$ as the IPR, and scaling functions $g_\eta$, $g_N$ depending
only on $W$. In fact, we expect more: the bulk correction is in both cases
the zero mode of the same linearized $\eta = 0$, $N = \infty$ cavity recursion,
onto which any weak perturbation injected in the far tail --- a finite
regulator as well as the loop-induced terms of a finite graph (see below) --- must relax,
the nature of the perturbation entering only through the amplitude. The
two scaling functions should therefore coincide up to a constant once
expressed in corresponding variables, $g_N(u) = A \, u^{-3} g_\eta(1/u)$
with $A = c_N/c_\eta$ [Fig.~\ref{fig:collapse}(b)].
Figure~\ref{fig:collapse} confirms this picture: the rescaled
differences collapse onto a single master curve over several orders of
magnitude in $\eta$ (one and a half decades in $N$), and the finite-$\eta$
and finite-$N$ master curves coincide, with $A$ close to one (drifting slowly with $N$, a subleading effect~\cite{companion}) --- a far more
stringent test than the exponent match of Fig.~\ref{fig:main}(c): the
correspondence between the two corrections extends to the bulk of the
distribution of the eigenstate amplitudes.

On a finite RRG, exact diagonalization shows the real-space origin of
the corrections directly~\cite{companion}: the distribution of the distance
between the two peaks of the doubly-peaked eigenstates is the same at all
$N$ up to a cutoff at $d \approx \ln N/\ln k$, of the order of the typical
distance between two sites. The eigenstates that would hybridize with a
more distant partner remain localized on a single center, with a larger IPR, and this
deficit accounts for most of the decrease of $I_2(N)$ with $N$ at $W = 22$, as implied
by Eq.~\eqref{eq:sumrule}.

In the cavity language, the corrections come from loops. The cavity
equations~\eqref{eq:cavity}--\eqref{eq:Gii} become exact only in the strict
$N\to\infty$ limit, where any two neighbors of a given node, once that node
is removed, are connected solely through paths of infinite length. A
finite RRG is only locally tree-like: a node lies on $\sim k^L/(2N)$ loops
of length $L$, rare for $L \ll \ln N/\ln k$ but of order one at $L \sim
\ln N/\ln k$, through which the two neighbors remain connected: these generate extra off-diagonal terms
$G_{m_1 m_2 \to i}$ on the right-hand side of Eq.~\eqref{eq:cavity},
responsible for the finite-$N$ corrections to the IPR.

We restore these terms in an approximate, self-consistent one-loop scheme
(see End Matter for details), in the spirit of the loop corrections to the
Bethe approximation for models on random
graphs~\cite{montanari2005compute,parisi2006loop,chertkov2006loop}, which
also control their finite-size corrections~\cite{ferrari2013finite,lucibello2014finite,metz2014finite}:
the $G_{m_1 m_2 \to i}$ terms are approximated by the contribution of the
\emph{shortest} path connecting $m_1$ and $m_2$ [Fig.~\ref{fig:sketch}(a)], which closes a loop through $i$, expressed as a product
of diagonal cavity Green's functions along that path~\cite{bandeira2026andersonlocalizationbethelattice}, with the path
length drawn from the loop statistics of the RRG~\cite{wormald1999models,marinari2004circuits,marinari2006number}. This closes the cavity recursion within an extended population dynamics.
Solving it for a given cutoff $\ell_{\max}$
on the path length (loop length $\ell+2$), mapped
onto an effective $N$ via $\ln N \simeq (\ell_{\max}+1)\ln k$, reproduces
the finite-$N$ correction to the IPR in quantitative agreement with exact diagonalization and its power-law extrapolation, in the regime where the scheme is under control [Fig.~\ref{fig:oneloop}].
Since the loop terms have a random sign, they act not as an average
regulator but as a rare-event perturbation of the far tail of
Eq.~\eqref{eq:betatail}, which excites the same zero mode as a finite
$\eta$ (see End Matter).

Why the cutoff of the resonances on a finite graph sets in at
splittings of the order of the mean level spacing, and how it relates to
the long loops of the one-loop scheme, remain open questions.

In summary, the corrections to the IPR in the localized phase of the
Anderson model on the Bethe lattice display a highly unconventional degree
of non-universality: both the finite-size and the finite-regulator
corrections decay as power laws governed by a single, continuously varying
exponent $\kappa(W) = 2\beta(W) - 1$, throughout the localized phase and
not merely near $W_c$. We derived this relation in a solvable toy NL$\sigma$M
sharing the symmetry of the true Anderson transition on the Bethe
lattice~\cite{zirnbauer1990fourier,zirnbauer1991fourier,Gruzberg_1996,dupre1996localization},
extended it to the cavity equations of the Anderson model, and confirmed it
numerically down to the full distributions and by a one-loop treatment of
the finite graphs; the complete analysis,
including the critical point, is given in Ref.~\cite{companion}.
Physically, finite $N$ and finite $\eta$ impose the same infrared cutoff on
the correlations of eigenstates close in energy, which are dominated by
rare Mott-like resonances between distant localized states: the rare
events in the tail of the LDoS distribution and these resonances are two
faces of the same phenomenon, $\beta > 1/2$ is the condition for them not
to proliferate, and on a finite graph the most distant hybridized pairs are
missing. Three consequences follow. Extrapolations of the IPR should use the scaling variable $(2\beta-1)\ln N$, whose coefficient varies throughout the localized phase and vanishes at $W_c$, where the correction is expected to become $\propto 1/\ln N$~\cite{companion}. This scaling form also fixes exactly the renormalization-group flow of the running fractal dimension $D_2 = -{\rm d}\ln I_2/{\rm d}\ln N$ proposed phenomenologically in Ref.~\cite{vanoni2023renormalization}: the flow is exactly of the Kosterlitz type~\cite{kosterlitz1974critical}, with $\kappa = 2\beta - 1$ conserved along each trajectory, as the invariant of the Kosterlitz equations, so that it can be integrated exactly, and each trajectory ends on a line of fixed points with ${\rm d}\ln D_2/{\rm d}\ln N = -(2\beta - 1)$, the same for all the running dimensions $D_q$ with $q > \beta$, whose corrections share the same scaling function~\cite{companion} (see also Ref.~\cite{scardicchio2026conformality}). Finally,
since the corrections come from the loops that the Bethe-lattice limit
discards, the one-loop scheme opens a route to a $1/d$ expansion around the
Bethe-lattice mean field, for Euclidean lattices and for MBL on
Hilbert-space graphs.

\begin{acknowledgments}
We thank W.~Chen, V.~E.~Kravtsov, G.~Lemari\'e, A.~D.~Mirlin, A.~Scardicchio and C.~Vanoni for invaluable discussions. We acknowledge financial support from the ANR research grant ManyBodyNet ANR-24-CE30-5851.
\end{acknowledgments}

\section*{End Matter}

\begin{figure*}[t]
\centering
\panel{a}{\includegraphics[width=0.325\textwidth]{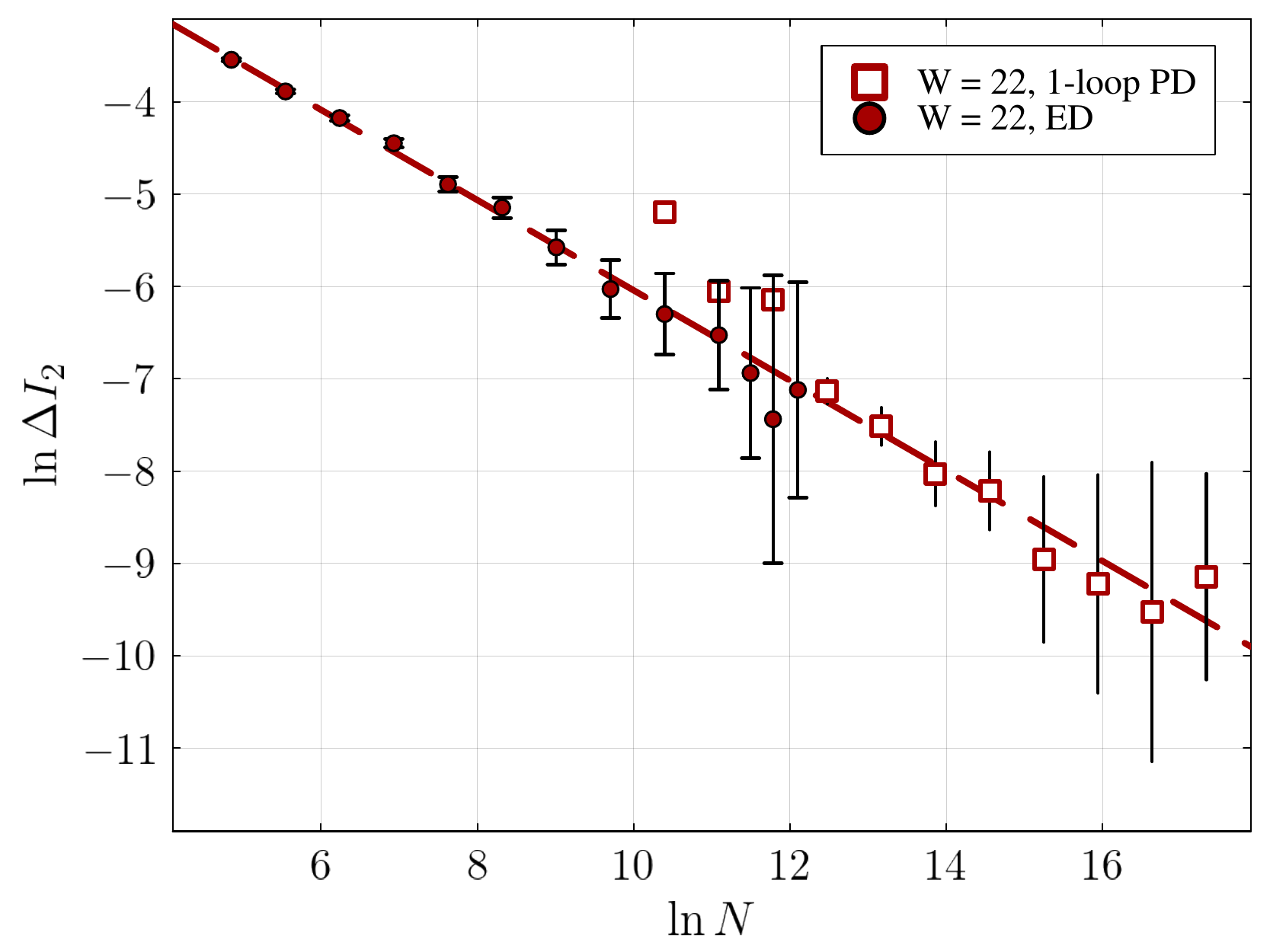}} \hfill
\panel{b}{\includegraphics[width=0.325\textwidth]{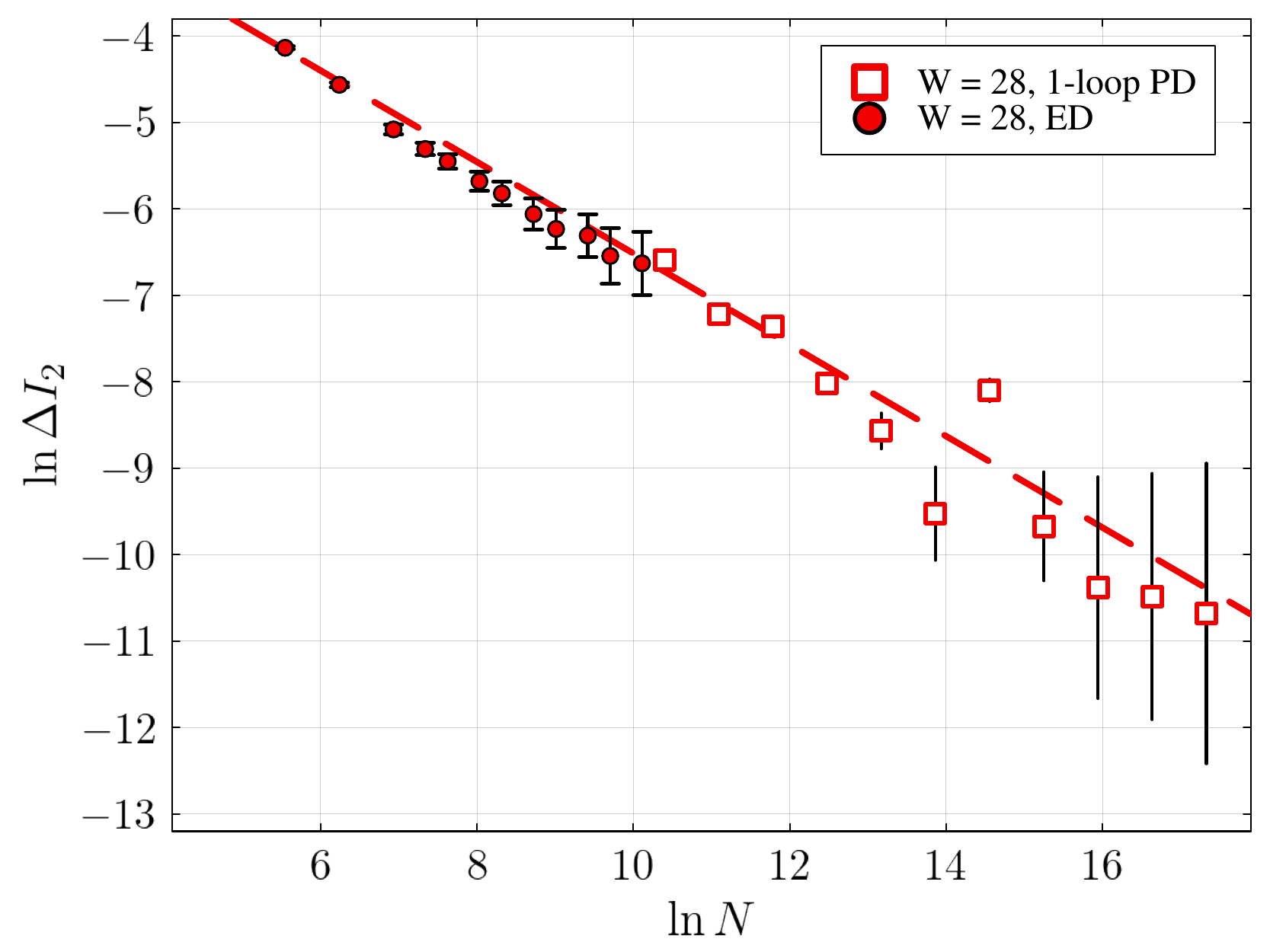}} \hfill
\panel{c}{\includegraphics[width=0.325\textwidth]{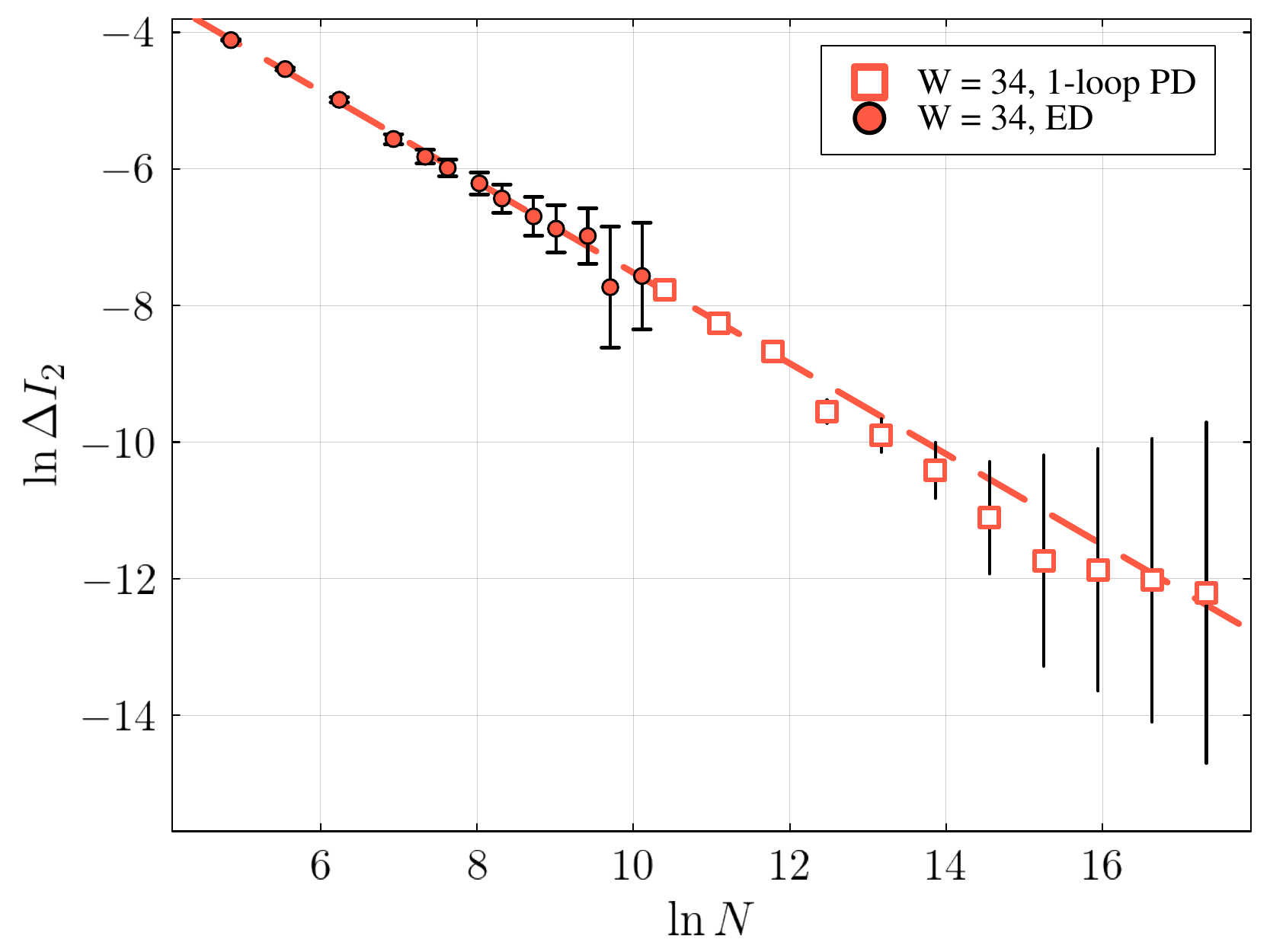}}
\caption{Finite-$N$ correction to the IPR, $I_2(N) - I_2^\infty$, as a
function of $N$ for $W=22$ (a), $W=28$ (b) and $W=34$ (c).
Filled symbols: exact diagonalization of the Anderson model on RRGs.
Empty symbols: self-consistent
one-loop population dynamics with cutoff $\ell_{\max}$, plotted at the
effective size $\ln N = (\ell_{\max}+1)\ln k$. Dashed lines: power-law
fits to Eq.~\eqref{eq:corrN}. The one-loop scheme reproduces the
exact-diagonalization results quantitatively, with no adjustable
parameters beyond the convention relating $\ell_{\max}$ to $N$; the residual discrepancy at $W=22$ and $\ell_{\max} \lesssim 16$
coincides with the regime in which the fraction of unphysical updates is
no longer negligible.}
\label{fig:oneloop}
\end{figure*}

The cavity equations~\eqref{eq:cavity}--\eqref{eq:Gii} are the $N \to \infty$
limit of a set of relations that are exact on {\it any} graph. Applying the
Schur complement formula to the resolvent of $\mathcal{H}$ with the row and column of
node $i$ removed, one obtains (with $t=1$)~\cite{bandeira2026andersonlocalizationbethelattice}
\begin{subequations} \label{eq:general}
\begin{align}
    G_{ii}^{-1} &= \epsilon_i - {\rm i}\eta - \!\! \sum_{m \in \partial i} \!\! G_{m \to i}
    - \!\!\!\! \sum_{m_1 \neq m_2 \in \partial i} \!\!\!\! G_{m_1 m_2 \to i} \, , \label{eq:generalGii} \\
    G_{i \to j}^{-1} &= \epsilon_i - {\rm i}\eta - \!\!\! \sum_{m \in \partial i \setminus j} \!\!\! G_{m \to i}
    - \!\!\!\!\! \sum_{m_1 \neq m_2 \in \partial i \setminus j} \!\!\!\!\! G_{m_1 m_2 \to i} \, , \label{eq:generalcav}
\end{align}
\end{subequations}
where $G_{m_1 m_2 \to i}$ are the {\it off-diagonal} elements of the resolvent
of the graph from which $i$ (and $j$) has been removed, and $G_{m \to i}
\equiv G_{mm \to i}$. In the absence of $i$, two of its neighbors $m_1$ and
$m_2$ remain connected only through the loops passing through $i$: on a
tree $G_{m_1 m_2 \to i} = 0$ and Eqs.~\eqref{eq:general} reduce to
Eqs.~\eqref{eq:cavity}--\eqref{eq:Gii}, while on a finite RRG the
off-diagonal terms are small but non-zero, and Eqs.~\eqref{eq:general} are
not closed.

\begin{figure}[t]
\centering
\begin{tikzpicture}[scale=0.8, every node/.style={font=\footnotesize},
  site/.style={circle,draw,fill=white,inner sep=1.6pt},
  cav/.style={circle,draw,dashed,fill=white,inner sep=1.6pt},
  ext/.style={circle,fill=black!60,inner sep=1.0pt},
  branch/.style={black!60,thin}]
\node[cav] (i) at (0,1.6) {};
\node[right=2pt of i] {$i$};
\node[cav] (j) at (0,2.9) {};
\node[right=2pt of j] {$j$};
\draw[dashed] (i)--(j);
\node[site] (m1) at (-1.6,0.6) {};
\node[site] (m2) at (1.6,0.6) {};
\draw[dashed] (i)--(m1); \draw[dashed] (i)--(m2);
\node[site] (x1) at (-1.6,-0.7) {};
\node[site] (x2) at (-0.6,-1.5) {};
\node[site] (x3) at (1.6,-0.7) {};
\draw[thick] (m1)--(x1)--(x2);
\draw[thick,dotted] (x2)--(0.6,-1.5);
\draw[thick] (0.6,-1.5)--(x3)--(m2);
\node at (-1.15,0.6) {$m_1$};
\node at (1.15,0.6) {$m_2$};
\node at (-1.2,-0.7) {$x_1$};
\node at (-0.45,-1.05) {$x_2$};
\node at (1.05,-0.7) {$x_{\ell-1}$};
\foreach \p/\d in {m1/170, x1/200, x2/265, x3/-20, m2/10} {
  \draw[branch] (\p) -- ++(\d:0.5) node[ext] {};
  \draw[branch] ($(\p)+(\d:0.5)$) -- ++(\d+25:0.32);
  \draw[branch] ($(\p)+(\d:0.5)$) -- ++(\d-25:0.32);
}
\node at (-2.5,2.9) {(a)};
\begin{scope}[xshift=5.4cm]
\node[cav] (ib) at (0,1.6) {};
\node[right=2pt of ib] {$i$};
\node[site] (m1b) at (-1.6,0.6) {};
\node[site] (m2b) at (1.6,0.6) {};
\node[site] (q2) at (1.6,1.7) {};
\node[right=2pt of q2] {$q_2$};
\draw[dashed] (ib)--(m1b);
\draw[thick] (m2b)--(q2);
\node[site] (x1b) at (-1.6,-0.7) {};
\node[site] (x2b) at (-0.6,-1.5) {};
\node[site] (x3b) at (1.6,-0.7) {};
\draw[thick] (m1b)--(x1b)--(x2b);
\draw[thick,dotted] (x2b)--(0.6,-1.5);
\draw[thick] (0.6,-1.5)--(x3b)--(m2b);
\node at (-1.15,0.6) {$m_1$};
\node at (1.15,0.6) {$m_2$};
\foreach \p/\d in {m1b/170, x1b/200, x2b/265, x3b/-20, m2b/10} {
  \draw[branch] (\p) -- ++(\d:0.5) node[ext] {};
  \draw[branch] ($(\p)+(\d:0.5)$) -- ++(\d+25:0.32);
  \draw[branch] ($(\p)+(\d:0.5)$) -- ++(\d-25:0.32);
}
\draw[-latex,thick,black!60] (1.05,-0.1) .. controls (0.6,-0.9) and (-0.4,-0.9) .. (-1.05,-0.1);
\node at (-2.5,2.9) {(b)};
\end{scope}
\end{tikzpicture}
\caption{Sketch of the one-loop construction for $k=2$. (a) In the absence
of $i$ (dashed), its neighbors $m_1$ and $m_2$ are connected by a shortest
path of $\ell$ edges through the sites $x_1, \ldots, x_{\ell-1}$, closing a
loop of length $\ell+2$ through $i$; every site on the path carries one
further branch (grey), described by a cavity Green's function drawn from
the population. The off-diagonal term $G_{m_1 m_2 \to i}$ is evaluated as
the product of diagonal cavity Green's functions along the path,
Eq.~\eqref{eq:offdiag}. (b) The diagonal term $G_{m_1 \to i}$ is computed
on the same realization by iterating the tree recursion along the open
chain from $m_2$ to $m_1$ (arrow), with $i$ replaced, as a neighbor of $m_2$,
by a fictitious node $q_2$ carrying a cavity Green's function from the
population, so that the cavity Green's function of $m_2$ that starts the
chain has $k$ incoming branches, as in the population; $G_{m_2 \to i}$ is
obtained symmetrically. This preserves the correlations between the
diagonal and off-diagonal terms that exist on a real graph.}
\label{fig:sketch}
\end{figure}
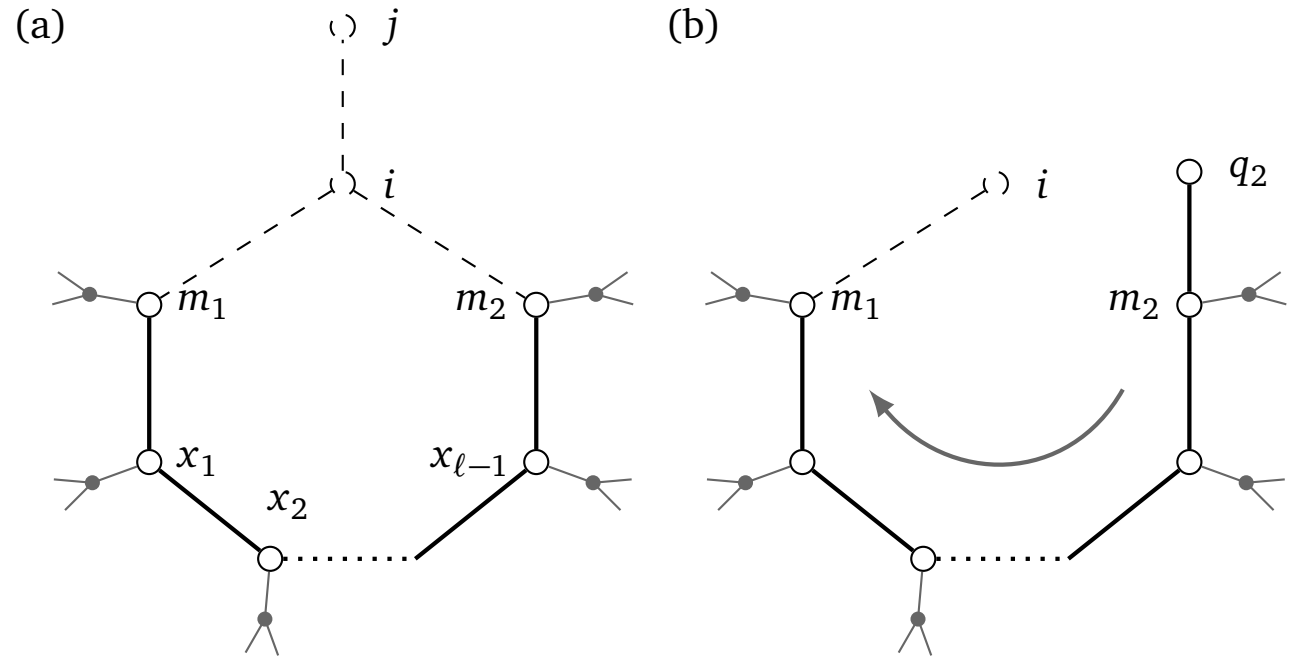

We close them by a one-loop approximation. Let $\ell$ be the length (number
of edges) of the shortest path connecting $m_1$ and $m_2$ in the absence of
$i$, running through the sites $x_0 \equiv m_1, x_1, \ldots, x_{\ell-1},
x_\ell \equiv m_2$ [Fig.~\ref{fig:sketch}(a)]. Retaining only this path and
treating it as embedded in an otherwise loopless tree, the off-diagonal
element factorizes into diagonal cavity Green's functions along the
path~\cite{bandeira2026andersonlocalizationbethelattice},
\begin{equation} \label{eq:offdiag}
    G_{m_1 m_2 \to i} = G_{m_1 \to i} \, \prod_{a=1}^{\ell} G_{x_a \to x_{a-1}} \, ,
\end{equation}
each factor being obtained from the standard tree recursion~\eqref{eq:cavity}
in terms of the on-site energies of the path and of the cavity Green's
functions of the branches attached to it.

The structure of the loop term is most transparent in the variables
appropriate to the $\eta \to 0^+$ limit of the localized phase, in which
the recursion is linearized in the imaginary parts~\cite{rizzo2024localized}.
Writing each of the $\ell+1$ factors in Eq.~\eqref{eq:offdiag} as
$G_{x_a \to x_{a-1}} = g_a + {\rm i}\eta\, \hat g_a + O(\eta^2)$ (with
$a=0$ labeling $G_{m_1 \to i}$, $g_a$ real and $\hat g_a > 0$) and
$\mathcal{G}_\ell \equiv \prod_{a=0}^{\ell} g_a$, one finds
${\rm Re}\, G_{m_1 m_2 \to i} = \mathcal{G}_\ell$ and
${\rm Im}\, G_{m_1 m_2 \to i} = \eta\, \mathcal{G}_\ell \sum_a \hat g_a/g_a$,
so that, in terms of $G_{i \to j}^{-1} = m_{i\to j} - {\rm i}\eta\, \hat m_{i\to j}$ and of
the neighbors' $G_{m \to i} = g_m + {\rm i}\eta\, \hat g_m$, the cavity
equation~\eqref{eq:generalcav} reads
\begin{equation} \label{eq:loopmhat}
\begin{aligned}
    m_{i\to j} &= \epsilon_i - \!\!\! \sum_{m \in \partial i \setminus j} \!\!\! g_m - 2\, \mathcal{G}_\ell \, , \\
    \hat m_{i\to j} &= 1 + \!\!\! \sum_{m \in \partial i \setminus j} \!\!\! \hat g_m
    + \sum_{a=0}^{\ell} \frac{2\, \mathcal{G}_\ell}{g_a} \, \hat g_a \, .
\end{aligned}
\end{equation}
The loop thus shifts the real part by $2\mathcal{G}_\ell$, the amplitude
transported once around the loop, and adds to the linear recursion for the
rescaled imaginary part $\hat m$ a term with exactly the same structure as
the tree terms: every site $a$ of the loop acts as an additional neighbor
of $i$, coupled with the coefficient $2\mathcal{G}_\ell/g_a = 2\prod_{a' \neq a}
g_{a'}$, i.e.\ the product of the amplitudes transported from $a$ to $i$ along the
two arcs of the loop. $\mathcal{G}_\ell$ is a product of $\ell+1$ real cavity Green's
functions of random sign, with a typical magnitude decaying exponentially
in $\ell$ and a broad distribution generated by the resonances along the
path. Because of its random sign, the loop term in $\hat m_{i\to j}$ can have
either sign, which is the origin of the occasional negative values of
$\hat m_{i\to j}$ discussed below.

The length $\ell$ is a random
variable whose statistics is that of the RRG: since the number of loops of
length $L$ through a given node is Poisson distributed with mean
$k^{L}/(2N)$~\cite{wormald1999models,marinari2004circuits,marinari2006number},
we draw $\ell$ from
\begin{equation} \label{eq:pl}
    p(\ell) \propto k^{\ell} \, , \qquad \ell = 1, \ldots, \ell_{\max} \, ,
\end{equation}
($\ell=1$ being a triangle through $i$). The cutoff $\ell_{\max}$ plays
the role of the graph diameter: it is fixed by requiring that every node
belongs, on average, to one loop of the maximal length $\ell_{\max}+2$,
$k^{\ell_{\max}+2}/(2N) = 1$, which for $k=2$ gives $\ln N =
(\ell_{\max}+1) \ln k$. This is the only place where $N$ enters the scheme.

The resulting ``one-loop improved'' population dynamics proceeds as
follows (we describe the case $k=2$). The population represents the
distribution of the cavity Green's functions. To generate a new element
$G_{i \to j}$: (1) draw $\epsilon_i$ uniformly in $[-W/2,W/2]$ and $\ell$ from $p(\ell)$;
(2) draw the on-site energies of $m_1, m_2$ and of the $\ell-1$ path sites,
and, for each of these $\ell+1$ sites, one cavity Green's function from the
population for its branch not belonging to the loop; (3) compute $G_{m_1 \to i}$
and $G_{m_2 \to i}$ {\it on the same realization}: on a real graph these
diagonal cavity Green's functions and the off-diagonal term all involve the
same loop and are strongly correlated. To preserve these correlations,
$G_{m_1 \to i}$ is obtained by starting from $m_2$ and iterating the tree
recursion site by site along the path up to $m_1$
[Fig.~\ref{fig:sketch}(b)]; in this open chain the removed site $i$ is
replaced, as a neighbor of $m_2$, by a fictitious node $q_2$ carrying a
cavity Green's function drawn from the population, so that every site keeps
the RRG connectivity. $G_{m_2 \to i}$ is obtained in the same way, starting
from $m_1$ with a fictitious node $q_1$ in place of $i$. The procedure is not
exact, since the true graph contains no such nodes, but it preserves as far
as possible the correlations between the diagonal and off-diagonal terms; (4) compute $G_{m_1 m_2 \to i}$ from Eq.~\eqref{eq:offdiag}, with the factors $G_{x_a \to x_{a-1}}$ and $G_{m_1 \to i}$ of the same chain; (5) insert everything in Eq.~\eqref{eq:generalcav},
$G_{i \to j}^{-1} = \epsilon_i - {\rm i}\eta - G_{m_1 \to i} - G_{m_2 \to i}
- 2 G_{m_1 m_2 \to i}$, and replace a randomly chosen element of the
population with the result. Because of the approximations involved
(shortest path only, stylized $p(\ell)$), the new element may occasionally
have a negative imaginary part, which is unphysical; in that case steps
(2)--(5) are repeated with the same $\ell$ (so as not to bias $p(\ell)$
against short loops, which produce the largest corrections) until a
physical value is obtained. The fraction of such events decreases
exponentially with $\ell_{\max}$ and with $W$, and the scheme is
considered under control when it is smaller than the relative correction
$[I_2(N) - I_2^\infty]/I_2 \lesssim 10^{-3}$ one aims to resolve; this restricts it to
$\ell_{\max} > 16$ at the smallest disorder shown in
Fig.~\ref{fig:oneloop}. Once the population is stationary, the local
Green's function~\eqref{eq:generalGii} is sampled in the same way, with
three neighbors and three independent path lengths $\ell_{12}, \ell_{13},
\ell_{23}$ drawn from $p(\ell)$, and $I_2$ follows from
Eq.~\eqref{eq:I2G}; in practice we work with the equations linearized in
the imaginary parts, Eq.~\eqref{eq:loopmhat}, as appropriate for $\eta \to 0^+$ in the localized
phase~\cite{rizzo2024localized}.

Figure~\ref{fig:oneloop} compares the finite-$N$ corrections obtained in
this way with exact diagonalization, for three disorder strengths in the
localized phase. The agreement is quantitative --- both the power-law
exponent $\kappa(W)$ and the prefactor are reproduced, the latter up to the
convention relating $\ell_{\max}$ to $N$, a change of which by a factor of
order one would shift the one-loop data by $O(1)$ in $\ln N$ and their
prefactor by a factor $O(1)^{\kappa}$, leaving the exponent unchanged ---
although the scheme contains no information about the actual graph beyond
the loop statistics~\eqref{eq:pl}. Since the one-loop data under control lie
mostly at sizes beyond those accessible to exact diagonalization, the
comparison is largely with the power-law extrapolation of the latter. This supports the interpretation given in the
main text: the finite-$N$ corrections to the IPR are quantitatively captured by
the one-loop terms of Eq.~\eqref{eq:loopmhat}. Two remarks qualify
this interpretation. First, short loops cannot produce the anomalous
correction: a loop of fixed length $\ell = O(1)$ transports an $O(1)$
amplitude, but is present at a given node with probability $k^{\ell+2}/(2N)$, so
that it only contributes a regular $O(1/N)$ correction, subleading with
respect to $N^{-\kappa}$ with $\kappa < 1$; indeed, retaining only
$\ell \le 2$ in the scheme yields corrections much smaller than the observed
ones. The anomalous correction comes from $\ell \simeq \ell_{\max}$, where
loops are present at every node but transport a small and broadly
distributed amplitude: it is a rare-event effect. Second, since
$\mathcal{G}_\ell$ has a random sign, the loop terms in
Eq.~\eqref{eq:loopmhat} do not renormalize $\eta$ or the on-site disorder
on average; being typically exponentially small in $\ell_{\max}$, they perturb the $\eta = 0$, $N = \infty$ recursion for $\hat m$
only in its far tail, at a scale $\hat m_N$ that depends on $N$ only through
$\ell_{\max}$. Under the assumptions of Ref.~\cite{companion}, the response of the fixed
point to a weak perturbation injected in its far tail is
universal: whatever its microscopic nature, the
perturbation relaxes onto the zero mode of the linearized recursion, with
tail $\propto \hat m^{-(2-\beta)}$, and matching the zero mode to the
unperturbed tail $\hat m^{-1-\beta}$ at the injection scale fixes its
amplitude to $\propto \hat m_N^{-(2\beta-1)}$. The agreement of $\kappa_N$ with
$\kappa_\eta$ in Fig.~\ref{fig:main}(c), within the accuracy of the fits, therefore indicates $\hat m_N \propto
N$ (equivalently, eigenstate amplitudes $u = 1/\hat m \sim 1/N$): the loops
excite the same zero mode that a regulator $\eta \propto 1/N$, injected at
$\hat m_\eta \sim 1/\eta$, would. A derivation of the injection scale from
the statistics of $\mathcal{G}_\ell$ along long loops is left for future
work.

\bibliography{references}

\end{document}